\documentstyle[preprint,aps]{revtex}
\begin{document}
\title{Equilibrium orbit analysis in a free-electron laser with a coaxial 
wiggler}

\author {B. Maraghechi$^{a,b}$, B. Farrokhi$^a$, J. E. Willett$^c$, and U.-H. Hwang$^d$}

\address{
{\it $^a$Institute for Studies in Theoretical Physics and Mathematics,\\}
{\it P.O. Box 19395-5531, Tehran, Iran.\\}
{\it $^b$Department of Physics, Amir Kabir University, Tehran, Iran.\\}
{\it $^c$Department of Physics and Astronomy, University of \\}
{\it Missouri-Columbia, Columbia, Missouri 65211.\\}
{\it $^d$Physics Department, Korea University of Technology and Education,
Chunan, Choongnam 330-860, Korea}
}
\maketitle
PACs number(s): 41.60.cr, 52.75.Ms

\begin{abstract}
An analysis of single-electron orbits in combined coaxial wiggler and axial
guide magnetic fields is presented. Solutions of the equations of motion are
developed in a form convenient for computing orbital velocity components and
trajectories in the radially dependent wiggler. Simple analytical solutions 
are obtained in the radially-uniform-wiggler approximation and a formula for 
the derivative of the axial velocity $v_{\|}$ with respect to 
Lorentz factor $\gamma$ is derived. Results of numerical computations are 
presented and the characteristics of the equilibrium orbits are discussed. 
The third spatial harmonic of the coaxial wiggler field gives rise to group 
$III$ orbits which are characterized by a strong negative mass regime.
\end{abstract}

\section*{I. INTRODUCTION}

Most free-electron lasers employ a wiggler with either a helically symmetric
magnetic field generated by bifilar current windings or a linearly symmetric
magnetic field generated by alternating stacks of permanent magnets. A uniform
static guide magnetic field is also frequently employed. Single-particle orbits
in these helical and planar fields combined with an axial guide field have been
analyzed in detail and have played a role in the development of free-electron
lasers\cite{r1}.
Harmonics of gyroresonance for off-axis electrons caused by the radial
 variation  of the magnetic field of a helical wiggler is found by Chu and Lin
\cite{r6}.
 Recently the feasibility of using a coaxial wiggler in a free-electron
laser has been investigated. Freund {\it et~al.}\cite{r2,r3} studied the performance
of a coaxial hybrid iron wiggler consisting of a central rod and a coaxial
ring of alternating ferrite and dielectric spacers inserted in a uniform
static axial magnetic field. McDermott {\it et~al.}\cite{r4} proposed the use of a
wiggler consisting of a coaxial periodic permanent magnet and transmission
line. Coaxial devices offer the possibility of generating higher power than
conventional free-electron lasers and with a reduction in the beam energy
required to generate radiation of a given wavelength.

In the present paper, single-particle orbits in a coaxial wiggler are studied.
The wiggler magnetic field is radially dependent with the fundamental plus the
third spatial harmonic component and a uniform static axial magnetic field
present. In Sec. II the scalar equations of motion are introduced and reduced
to a form which is correct to first order in the wiggler field. In Sec. III
solutions of the equations of motion are developed in a form suitable for
computing the electron orbital velocity and trajectory in the radially
dependent magnetic field of a coaxial wiggler. The special case of a radially
independent wiggler is also analyzed. In Sec. IV the results of numerical
computations of the wiggler field components, velocity components, radial
excursions, and the $\Phi$ function for locating negative mass regimes are
presented and discussed. In Sec. V some conclusions are presented.

\section*{II. EQUATIONS OF MOTION}

Electron motions in a static magnetic field ${\bf B}$ may be determined by 
solution of the vector equation of motion
\begin{equation}
\frac{d{\bf v}}{dt}=\frac{-e}{\gamma mc}{\bf v}\times {\bf B}
\end{equation}
where ${\bf v}$, $-e$, and $m$ are the velocity, charge, and (rest) mass, 
respectively, of the electron. Lorentz factor $\gamma$ is a constant given by 
\begin{equation}
\gamma = (1-v^2/c^2)^{-1/2}
\end{equation}
where $v=|{\bf v}|$ is the constant electron speed.

The total magnetic field inside a coaxial wiggler will be taken to be of the 
form
\begin{eqnarray}
&&{\bf B}=B_r\hat {\bf r}+B_z\hat {\bf z},\\
&&B_r=B_wF_r(r,z),\\
&&B_z=B_0+B_wF_z(r,z),
\end{eqnarray}
where $B_0$ is a uniform static axial guide field, and $F_r$ and $F_z$ are 
known functions of cylindrical coordinates $r$ and $z$. Equation (1) may be 
written in the scalar form
\begin{eqnarray}
&&\frac{dv_r}{dt}-\frac{v_{\theta}^2}{r}=-v_{\theta}(\Omega_0+\Omega_wF_z),\\
&&\frac{dv_{\theta}}{dt}+\frac{v_{\theta}v_r}{r}=v_r(\Omega_0+\Omega_wF_z)-v_z\Omega_wF_r,\\
&&\frac{dv_z}{dt}=v_{\theta}\Omega_wF_r;
\end{eqnarray}
$\Omega_0$ and $\Omega_w$ are relativistic cyclotron frequencies given by
\begin{eqnarray}
&&\Omega_0=\frac{eB_0}{\gamma mc},\\
&&\Omega_w=\frac{eB_w}{\gamma mc}.
\end{eqnarray}

Initial conditions will be chosen such that the transverse motion of the 
electron in the $B_0$ field vanishes in the limit as $B_w$ approaches zero. 
Then, in order to develop a solution to first order in the wiggler field 
$B_w$, the scalar equations of motion will be approximated by 
\begin{eqnarray}
&&\frac{dv_r}{dt}=-\Omega_0v_{\theta},\\
&&\frac{dv_{\theta}}{dt}=\Omega_0v_r-v_{\Vert}\Omega_wF_r,\\
&&\frac{dv_z}{dt}=0.
\end{eqnarray}
with the wiggler field approximated by the fundamental plus the third spatial 
harmonic component,
\begin{equation}
F_r=F_{r1}sin(k_wz)+F_{r3}sin(3k_wz),
\end{equation}
where
\begin{eqnarray}
&&F_{rn}\equiv G_n^{-1}[S_nI_1(nk_wr)+T_nK_1(nk_wr)],\\
&&G_n\equiv I_0(nk_wR_{out})K_0(nk_wR_{in})-I_0(nk_wR_{in})K_0(nk_wR_{out}),\\
&&S_n\equiv \frac{2}{n\pi}sin(\frac{n\pi}{2})[K_0(nk_wR_{in})+K_0(nk_wR_{out})],\\
&&T_n\equiv \frac{2}{n\pi}sin(\frac{n\pi}{2})[I_0(nk_wR_{in})+I_0(nk_wR_{out})],
\end{eqnarray}
and $n=1,3$; $R_{in}$ and $R_{out}$ are the inner and outer radii of the 
coaxial waveguide, $k_w=2\pi /\lambda_w$ where $\lambda_w$ is the wiggler
(spatial) period, and $I_0$, $I_1$, $K_0$, and $K_1$ are modified Bessel 
functions.

\section*{III. ORBITAL ANALYSIS}

{\bf A. Radially dependent wiggler}

The scalar equations of motion may be solved to determine the electron 
orbital velocity and trajectory in a coaxial wiggler. Equation (13) yields
\begin{equation}
v_z=v_{\Vert}
\end{equation}
where the constant $v_{\Vert}$ is the root-mean-square axial velocity 
component. With the initial axial position taken to be $z_0=0$,
\begin{equation}
z=v_{\Vert}t.
\end{equation}
Equations (11), (12), (14), and (20) may be combined to obtain
\begin{equation}
\frac{d^2v_r}{dt^2}+\Omega_0^2v_r=f(t)
\end{equation}
where
\begin{equation}
f(t)=\Omega_0 \Omega_wv_{\Vert}[F_{r1}sin(k_wv_{\Vert}t)+F_{r3}sin(3k_wv_{\Vert}t)].
\end{equation}
By the method of variation of parameters, a solution of Eq. (21) may be 
obtained in the form
\begin{eqnarray}
v_r=[-v_{\theta 0}+\Omega_0^{-1}\int_0^t\ f(\tau)cos(\Omega_0\tau)\ d\tau]sin(\Omega_0t)\nonumber\\
+[v_{r0}-\Omega_0^{-1}\int_0^t\ f(\tau)sin(\Omega_0\tau)\ d\tau]cos(\Omega_0t)
\end{eqnarray}
where $v_{r0}$ and $v_{\theta 0}$ are initial radial and azimuthal velocity 
components. Then Eq. (11) yields
\begin{eqnarray}
v_{\theta}=[v_{\theta 0}-\Omega_0^{-1}\int_0^t\ f(\tau)cos(\Omega_0\tau)\ d\tau]cos(\Omega_0t)\nonumber\\
+[v_{r0}-\Omega_0^{-1}\int_0^t\ f(\tau)sin(\Omega_0\tau)\ d\tau]sin(\Omega_0t).
\end{eqnarray}
The orbital velocity is given to first order in $B_w$ by Eqs. (23), (24), 
and (19). The trajectory may then be computed using
\begin{eqnarray}
&&r=r_0+\int_0^t\ v_r(\tau)d\tau,\\
&&\theta=\theta_0+\int_0^t\ v_{\theta}(\tau) d\tau,
\end{eqnarray}
and Eq. (20).

{\bf B. Radially uniform wiggler}

By neglecting the radial variation of $F_{r1}$ and $F_{r3}$, a solution of 
Eq. (21) may be obtained in the form
\begin{equation}
v_r=\alpha_1sin(k_wv_{\Vert}t)+\alpha_3sin(3k_wv_{\Vert}t),
\end{equation}
where
\begin{equation}
\alpha_n=\frac{\Omega_0 \Omega_wv_{\Vert}F_{rn}}{\Omega_0^2-n^2k_w^2v_{\Vert}^2} \hskip 1cm (n=1,3).
\end{equation}
Equation (11) then yields
\begin{equation}
v_{\theta}=-\Omega_0^{-1}k_wv_{\Vert}\alpha_1cos(k_wv_{\Vert}t)-\Omega_0^{-1}(3k_wv_{\Vert}\alpha_3)cos(3k_wv_{\Vert}t).
\end{equation}
The corresponding initial conditions are
\begin{equation}
v_{r0}=0,
\end{equation}
\begin{equation}
v_{\theta 0}=-\Omega_0^{-1}k_wv_{\Vert}\alpha_1-\Omega_0^{-1}(3k_wv_{\Vert}\alpha_3).
\end{equation}

Root-mean-square values of the velocity components may be determined by use of 
Eqs. (27), (28), and (19). Replacing $v^2$ by its root-mean-square value in 
Eq. (2) then yields
\begin{equation}
\frac{v_{\Vert}^2}{c^2}[1+\frac{1}{2}(\frac{\alpha_1}{v_{\Vert}})^2+\frac{1}{2}\Omega_0^{-2}k_w^2\alpha_1^2
+\frac{1}{2}(\frac{\alpha_3}{v_{\Vert}})^2+\frac{9}{2}\Omega_0^{-2}k_w^2\alpha_3^2]=1-\gamma^{-2}.
\end{equation}
The derivative of $v_{\Vert}$ with respect to $\gamma$ may be obtained from 
Eq. (32) and, after some algebra, cast into the form
\begin{equation}
\frac{dv_{\Vert}}{d\gamma}=\frac{c^2}{\gamma \gamma_{\Vert}^2v_{\Vert}}\Phi
\end{equation}
where
\begin{equation}
\Phi=1-\frac{\sum_{n=1,3}\ (\Omega_0^2-n^2k_w^2v_{\Vert}^2)^{-3}\gamma_{\Vert}^2\Omega_w^2F_{rn}^2\Omega_0^2
(\Omega_0^2+3n^2k_w^2v_{\Vert}^2)}{2+\sum_{n=1,3}\ (\Omega_0^2-n^2k_w^2v_{\Vert}^2)^{-3}\Omega_w^2F_{rn}^2\Omega_0^2(\Omega_0^2+3n^2k_w^2v_{\Vert}^2)}.
\end{equation}
This equation may be used to establish the existence of a negative mass regime.

\section*{IV. NUMERICAL RESULTS}

A numerical computation is conducted to investigate the properties of the
equilibrium orbits of electrons inside a coaxial wiggler. Wiggler wavelength
 $2\pi/k_w$ and lab-frame
electron density $n_0$ were taken to be 3 cm and $10^{12}$ cm$^{-3}$, respectively.
The wiggler magnetic field $B_w$ was
taken to be 3745 G which corresponds to the relativistic wiggler frequency
$\Omega_w/ck_w=0.442$. Electron-beam energy $(\gamma -1)m_0c^2$ was taken to
 be $700$ keV, corresponding to a Lorentz factor $\gamma =2.37$. The axial
magnetic field $B_0$ was varied from  0 to 25.3 kG, corresponding to a
variation from 0 to 3 in the normalized relativistic cyclotron frequency
$\Omega_0/ck_w$ associated with $B_0$. The inner and outer radii of the coaxial
wiggler were assumed to be $R_{in}=1.5$ cm and $R_{out}=3$ cm, respectively.

Figure 1 shows the variation of the axial velocity of the quasi-steady-state
orbits with the axial guide magnetic field for three classes of solutions.
Group I orbits for which $0<\Omega_0<k_wv_{\Vert}$, group II orbits with
$k_wv_{\Vert}<\Omega_0<3k_wv_{\Vert}$, and group III orbits with
$\Omega_0>3k_wv_{\Vert}$. Existence of group III orbits is due to the presence
of the third spatial harmonic of the wiggler field, which also produces the
second magnetoresonance at $\Omega_0\approx 3k_wv_{\Vert}$.
The narrow width of the second resonance at $\Omega_0/ck_w\approx 2.7$ compared
to the width of the first magnetoresonance at $\Omega_0\approx k_wv_{\Vert}$ is
illustrated in Fig. 1B.
This is due to the relatively weak third harmonic compared to the fundamental
component of the wiggler field.
It should be noted that although the exact resonances $\Omega_0=k_wv_{\Vert}$
and $\Omega_0=3k_wv_{\Vert}$ occur at the origin where
$v_{\Vert}/c=\Omega_0/ck_w=0$, the first "magnetoresonance" in the literature
refers to the group II orbits with the cyclotron frequencies around
$\Omega_0/ck_w\approx 1$ in Fig. 1. Similarly we refer to the group III orbits
with the cyclotron frequency around $\Omega_0/ck_w\approx 2.7$ as the second
magnetoresonance.

The rate of change of the electron axial velocity with electron energy is
proportional to  ${\it \Phi}$ and is equal to unity in the absence of the wiggler
field.
Figure 2 illustrates the dependence of ${\it \Phi}$ on the radial wiggler magnetic
field and the axial guide magnetic field $B_0$. The curves corresponding to
the group I and II orbits are almost unaffected by the  third harmonic and are
almost the same as in Ref. \cite{r5} where the third harmonic is neglected. A negative
mass regime (i.e., negative $\Phi$ for which a decrease in the axial velocity
results in   an increase in the electron energy) is found for group III
orbits which is stronger than that of the group II orbits.

Equations (15)-(18) are used to calculate the radial components of the
wiggler field $F_{r1}$ and $F_{r3}$. For the axial component the following
expressions are used \cite{r3}
\begin{equation}
F_z=F_{z1}\ cos(k_wz)+F_{z3}\ cos(3k_wz),
\end{equation}
\begin{equation}
F_{zn}=G_n^{-1}\ [S_nI_0(nk_wr)-T_nK_0(nk_wr)].
\end{equation}
Figure 3 shows the variation of the amplitudes of the wiggler magnetic field
(divided by $B_w=3745$ G) with radius, for the first and
third spatial harmonics.
For the first harmonic the radial component, $F_{r1}$, has a minimum at
$r\approx 2.28$ cm and the axial component, $F_{z1}$, changes sign around this
point. Mc Dermott ${\it et}$ ${\it al.}$ \cite{r4} have demonstrated the stability
of a thin annular electron beam when $F_{r1}$ is minimum at the  beam radius.
The radial and axial components of the third harmonic of the wiggler $F_{r3}$
 and $F_{z3}$ are also shown in Fig. 3. Magnitudes of $F_{r3}$ and $F_{z3}$
both are minimum at $r\approx 2.28$ cm
where $F_{r1}$ is minimum. This is actually an inflection point for $F_{z3}$.

Variations of the radial components of the  first and third spatial harmonics
of the normalized wiggler magnetic field $F_{r1}$ and $F_{r3}$ with the wiggler
wave number $k_w$ are shown in Fig. 4.
Figure 4 also shows the dimensionless transverse velocity coefficients
$\bar \alpha_1=\alpha_1/c$ and $\bar \alpha_3=\alpha_3/c$ for the initial
orbit radius $r_0\approx 2.28$ cm where $F_{r1}$ is minimum. The cyclotron
frequency $\Omega_0/ck_w\approx 2.7$ is taken at the second magnetoresonance,
and our choice  of 3 cm for the wiggler wavelength corresponds to
$k_w\approx 2.1$ cm$^{-1}$.
It can be observed that at this wave number although the radial component of
the wiggler field at the first harmonic $F_{r1}$ is much larger than the
third harmonic $F_{r3}$, the transverse velocity coefficients of the third
harmonic $\bar \alpha_3$ are larger than $\bar \alpha_1$. This shows that the
third harmonic may have considerable effects around the second magnetoresonance
at $\Omega_0 \approx 3k_wv_{\Vert}$. Away from this resonance Eq. (28) shows that
$\alpha_3$ will be of the order of $F_{r3}$.

In order to study the transverse motion of electrons in the radially
dependent wiggler field Eqs. (11), (12), (25), and (26) are solved numerically
with the initial conditions chosen so that, in the limit of zero wiggler field,
there is axial motion at constant velocity $v_{\Vert}$ but no Larmor motion.
Figure 5 shows the variation of the radial and azimuthal components of electron
velocity with $z~(=v_{\Vert}t)$.
The normalized cyclotron frequency $\Omega_0/ck_w$ is chosen to be 0.5, 1.2,
and 3 for group
I, II, and III orbits, respectively, which are somewhat away from the
magnetoresonances. Solid curves correspond to the initial orbit radius
$r_0=2.28$ cm, which is at the point where $F_{r1}$ is minimum. Broken curves
correspond to $r_0=1.8$ cm, which is away from the $F_{r1}$ minimum. It can be
observed that the spatial periodicity of $v_r$ and $v_{\theta}$ for the first
two groups is equal to one wiggler wavelength, which is the same as that of
the first harmonic. Although group III orbits have a clear sinusoidal shape at
$r_0=2.28$ cm (solid curves), the slight deviations from sinusoidal shape are
obvious at $r_0=1.8$ cm (broken curves). This is because that at $r_0=2.28$ cm
where $F_{r1}$ is minimum $F_{r3}$ is very small. Therefore away from the
resonance the third harmonic plays almost no role, at $r_0=2.28$ cm. Moving
away from $F_{r1}$ minimum to $r_0=1.8$ cm, however, increases the magnitude of
$F_{r3}$ slightly making the effect of the third harmonic noticeable on group
III orbits, which are away from the second magnetoresonance, in Fig. 5.

Figure 6 shows the variations of $v_r/c$ and $v_{\theta}/c$ with $z~(=v_{\Vert}t)$
for group III orbits when the cyclotron frequency is adjusted at the second
magnetoresonance at $\Omega_0/ck_w\approx 2.7$. At $r_0=2.28$ cm the periodicity
is approximately equal to $\lambda_w/3$, which is the same as that of the
third harmonic, and shows the strong influence of the third harmonic
on the transverse velocity components. Going
away from the $F_{r1}$ minimum to $r_0=1.8$ cm makes the amplitudes of oscillations
of $v_r$ and $v_{\theta}$ larger.
Broken curves correspond to the solutions of Eqs. (27) and (29) when
the radial variation of the wiggler field is neglected. These solutions do not
differ appreciably from the r-dependent solutions for $r_0=2.28$ cm because
at $F_{r1}$ minimum the radial excursions are small for group III orbits. Away
from the $F_{r1}$ minimum at $r_0=1.8$ cm, however, deviations are noticeable
due to the larger radial excursions.

Figure 7 shows $v_r/c$ versus $z$ for group II orbits for r-dependent wiggler
(broken curves) and r-independent wiggler (solid curves).
The initial orbit radius is taken at $r_0=2.28$ cm and
the cyclotron frequency
is chosen around the first magnetoresonance at $\Omega_0/ck_w=0.9$. Large radial
excursions of electrons for group II orbits make the transverse velocity strongly
affected by the radial dependency of the wiggler field. The amplitude is also
modulated in space with the wavelength of around $16\lambda_w=48$ cm.

The radial excursion $r$ shown in Fig. 8 corresponds to the cyclotron
frequencies away from the magnetoresonances at $\Omega_0/ck_w$ equal to
0.5, 1.2, and 3 for the group I, II, and III orbits, respectively. Solid curves
correspond to the $r_0=2.28$ cm and the broken curves correspond to
$r_0=1.8$ cm.
It can be noticed that when the electrons are injected into the wiggler at
$r_0=2.28$ cm, where $F_{r1}$ is minimum, electron orbits remain well away from
the waveguide walls at $R_{in}=1.5$ cm and $R_{out}=3$ cm.

Figure 9 compares the radial excursions of group III orbits at the second
magnetoresonance at $\Omega_0/ck_w=2.7$, (solid curve) with those slightly away
from the resonance at $\Omega_0/ck_w=3$, (broken curve). Influence of the third
harmonic can be clearly seen through the modulation of the third harmonic by
the first harmonic when the cyclotron frequency is adjusted at the second
magnetoresonance.

\section*{V. CONCLUSIONS}

The third spatial harmonic of the coaxial wiggler field gives rise to the
group III orbits with $\Omega_0>3k_wv_{\Vert}$. This relatively weak third
harmonic makes the width of the second magnetoresonance narrow compared to
the first magnetoresonance. A strong negative mass regime is found for the
group III orbits. By adjusting the cyclotron frequency at the second
magnetoresonance the wiggler induced velocity of the group III orbits was
found to increase considerably. When the electrons are injected into the
wiggler where its magnetic field is minimum the electron orbits remain
well away from the waveguide boundaries.

Harmonic gyroresonance of electrons in the combined helical wiggler
and axial guide magnetic field is reported by Chu and Lin \cite{r6}.
In their analysis the relativistic single particle equation of motion 
is used with the axial velocity as well as the axial magnetic field
 of the wiggler averaged along the axial direction.  By assuming near 
steady-state orbits for off-axis electrons they found that the radial
variation of the wiggler magnetic field produces a harmonic structure
 in the transverse force. This force, in turn, comprises oscillations 
at all harmonics of $k_wz$.  It should be noted that there is no 
harmonic structure in the helical wiggler itself and the higher velocity 
harmonics vanish for the exact steady-state orbits of the on-axis 
electrons.  Moreover, higher harmonics do not appear in the one 
dimensional helical wiggler where the radial variation is neglected.

In the present analysis of coaxial wiggler, on the other hand,
equation of motion is written to first order in the wiggler 
amplitude. With this approximation axial component of the wiggler 
field has no contribution to the problem leaving the axial
 velocity as a constant.
Magnetic field of a coaxial wiggler is composed of a fundamental
plus a large number of odd spatial harmonics, which directly appear
in the magnetic force represented by $f(t)$ in Eq. (22).  Third 
harmonic in $f(t)$ appears in the transverse velocity components 
as a part of the integrands in Eqs. (23) and (24) and is also
demonstrated numerically for the radially dependent coaxial wiggler, 
but for the radially uniform wiggler the
third harmonic is explicit in the solutions 
 Eqs. (27)-(29).

\begin{figure}
\caption
{Normalized axial velocity $v_{\Vert}/c$ as a function of the normalized
axial-guide magnetic field $\Omega_0/ck_w$ for group I, II, and III orbits.
Narrow width of the second resonance at $\Omega_0/ck_w\approx 2.7$ compared
to the width of the first magnetoresonance
 at $\Omega_0\approx k_wv_{\Vert}$ is
illustrated in Fig. 1B.}
\vskip 0.5cm
\caption
{Factor ${\it \Phi}$ as a function of the normalized axial-guide magnetic
field $\Omega_0/ck_w$ for group I, II, and III orbits.}
\vskip 0.5cm
\caption
{Radial dependence of the radial and axial magnetic fields (divided by
$B_w=3745$ G) in the coaxial wiggler for the fundamental and third spatial
harmonics.}
\vskip 0.5cm
\caption
{Wave-number dependence of the radial components of the normalized
wiggler magnetic field $F_{r1}$, $F_{r3}$ and the dimensionless transverse
velocity coefficients $\bar \alpha_1$, $\bar \alpha_3$. The normalized
cyclotron frequency $\Omega_0/ck_w=2.7$ is taken at the second
magnetoresonance and $r_0=2.28$ cm.}
\vskip 0.5cm
\caption
{Normalized transverse velocity components as a function of axial distance
$z$ for the initial orbit radius $r_0=2.28$ cm (solid curves) and $r_0=1.8$ cm
(dotted curves). The normalized cyclotron frequency $\Omega_0/ck_w$ is
0.5, 1.2, and 3 for the group I, II, and III orbits, respectively.}
\vskip 0.5cm
\caption
{Normalized transverse velocity components as a function of axial distance
$z$ for group III orbits, at the second magnetoresonance $\Omega_0/ck_w=2.7$, 
solid curves correspond to the radial dependent wiggler and broken curves
correspond to the radial independent wiggler.}
\vskip 0.5cm
\caption
{Normalized radial velocity as a function of axial distance $z$ for
group II orbits at the first magnetoresonance $\Omega_0/ck_w=0.9$ with
$r_0=2.28$ cm. Solid curves correspond to the radial dependent wiggler and
broken curves correspond to the radial independent wiggler.}
\vskip 0.5cm
\caption
{Radial excursion $r$ as a function of axial distance $z$
for $r_0=2.28$ cm
(solid curves) and $r_0=1.8$ cm (broken curves) for groups I, II, and III
orbits out of resonance at $\Omega_0/ck_w=$0.5, 1.2, and 3, respectively.}
\vskip 0.5cm
\caption
{Radial excursion $r$ as a function of axial distance $z$
for group III orbits at $r_0=2.28$ cm. Solid curve corresponds to resonance
at $\Omega_0/ck_w=2.7$ and broken curve corresponds to out of resonance
at $\Omega_0/ck_w=3$.}
\end{figure}
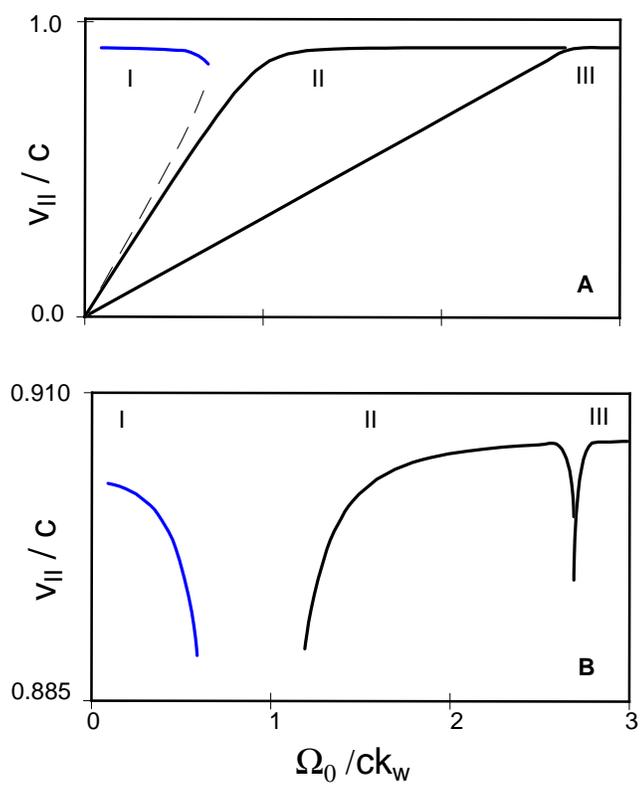
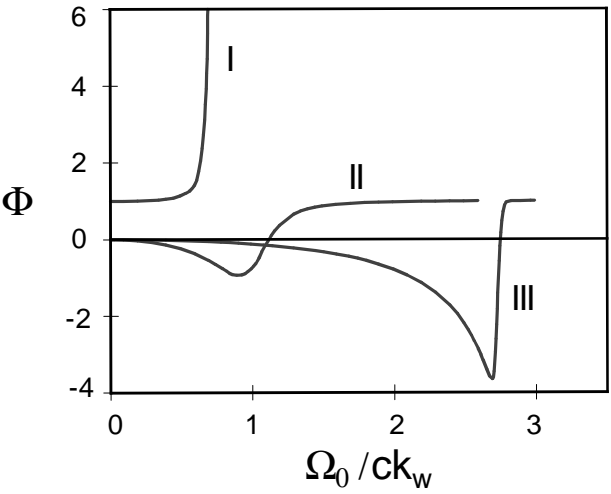
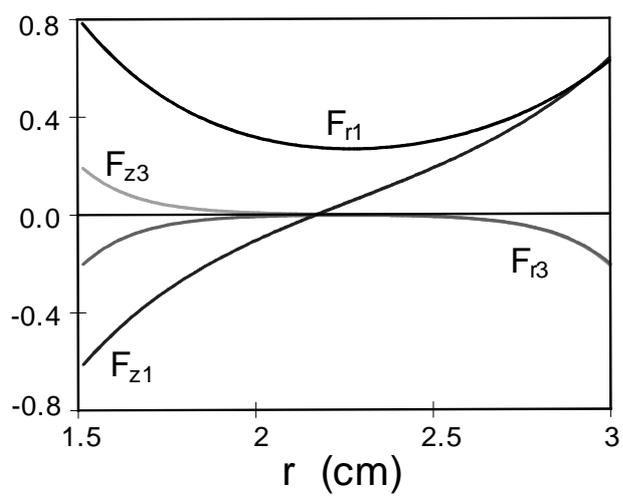
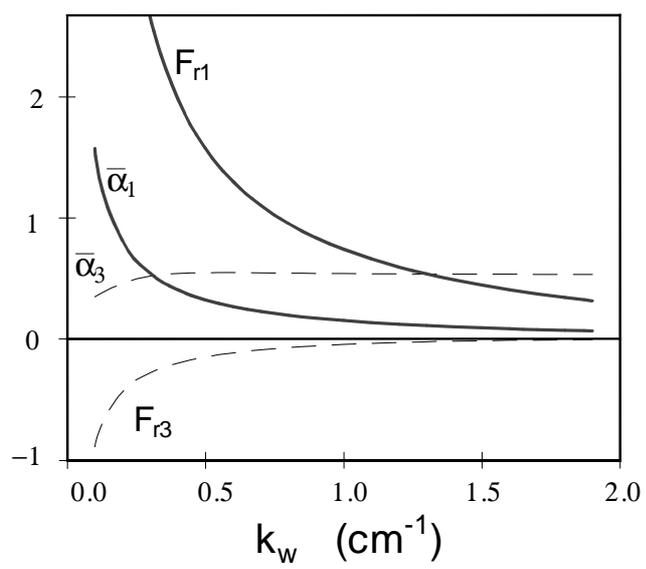
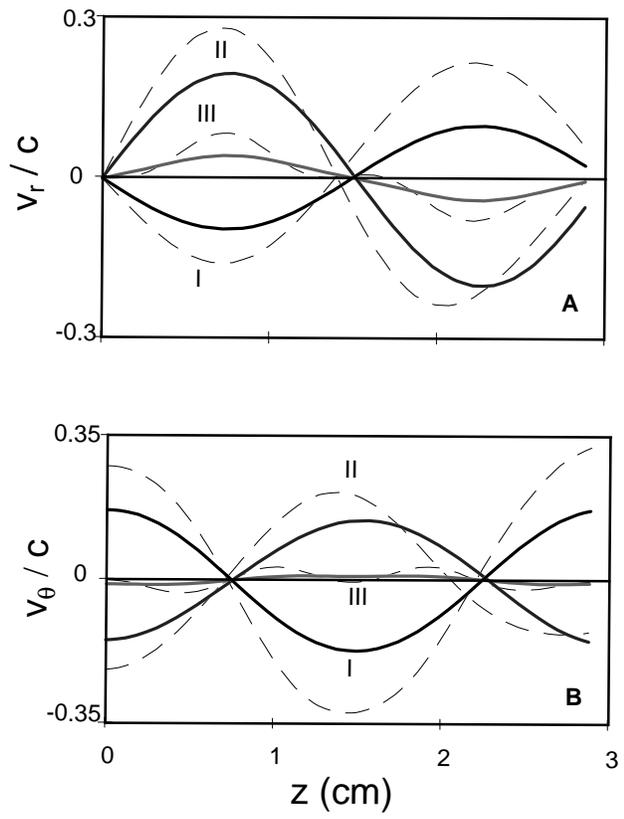
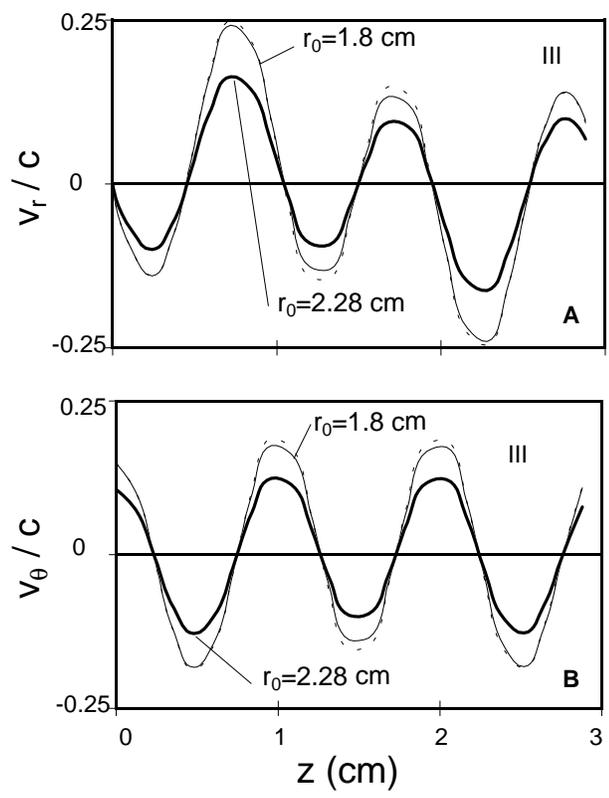
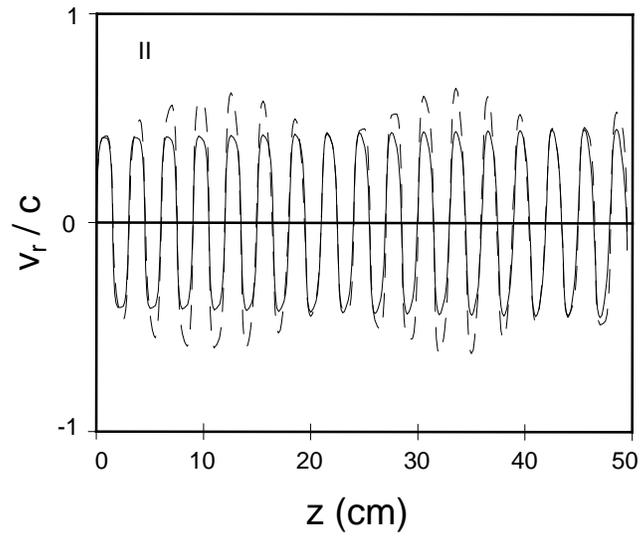
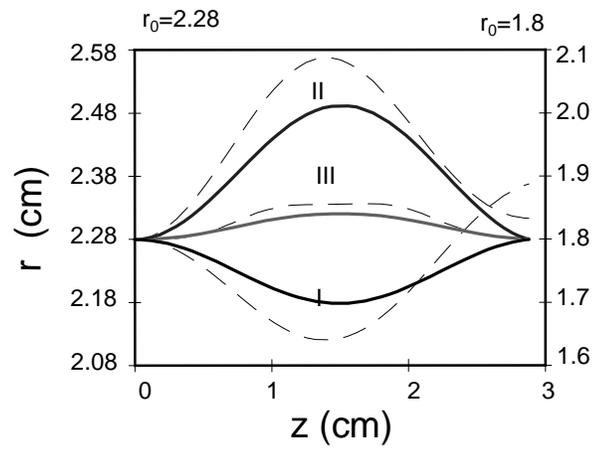
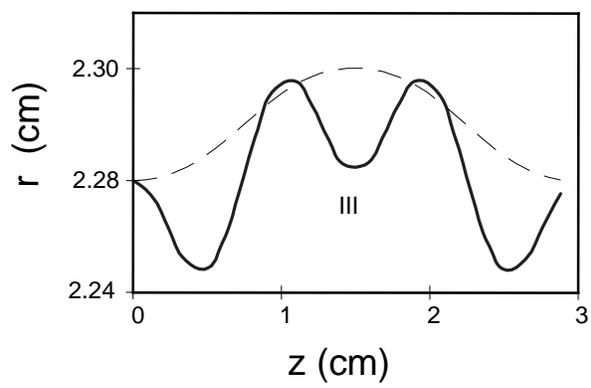
\end{document}